\documentclass[aps,pra,superscriptaddress,twocolumn]{revtex4-1}
\usepackage{bm} \usepackage{graphicx} \usepackage{amsmath}
\usepackage{amsthm,stmaryrd,mathtools}
\usepackage{amssymb} \usepackage{epstopdf} \usepackage{txfonts}

\newcommand{\prlsection}[1]{{\em {#1}.--~}}

\DeclareMathOperator*{\Tr}{Tr}

\DeclareRobustCommand\openzero{\leavevmode\hbox{0\kern-.55em0}}
\mathchardef\minus="002D

\newcommand{\ket}[1]{|{#1}\rangle}
\newcommand{\bra}[1]{\langle{#1}|}
\newcommand{\mean}[1]{\langle{#1}\rangle}

\newcommand\ua\uparrow
\newcommand\da\downarrow

\begin{document}

\title{
  Entanglement entropy scaling in solid-state spin arrays via 
  capacitance measurements
}

\author{Leonardo Banchi}
\affiliation{Department of Physics and Astronomy, University College London, Gower Street, WC1E 6BT London, United Kingdom}

\author{Abolfazl Bayat}
\affiliation{Department of Physics and Astronomy, University College London, Gower Street, WC1E 6BT London, United Kingdom}

\author{Sougato Bose}
\affiliation{Department of Physics and Astronomy, University College London, Gower Street, WC1E 6BT London, United Kingdom}

\date{\today}

\begin{abstract}
  Solid-state spin arrays are being engineered in varied systems,
  including gated coupled quantum dots and interacting dopants in semiconductor 
  structures.
  Beyond quantum computation, these arrays are useful integrated  analog simulators
  for many-body models. 
  As entanglement between individual spins is extremely short ranged 
  in these models, 
  one has to measure the entanglement entropy 
  of a block in order to truly
  verify their many-body entangled nature. 
  Remarkably, the characteristic scaling of entanglement entropy, 
  predicted by conformal field theory, has never been measured. 
  Here we show that with as few as two  
  replicas  of a spin array, and capacitive double-dot singlet-triplet measurements on 
  neighboring spin pairs, the above scaling of the entanglement entropy can be 
  verified. This opens up the controlled simulation of quantum field theories, 
  as we exemplify with uniform chains and Kondo-type impurity models, 
  in engineered solid-state systems.  Our 
  procedure remains effective even in the presence of typical imperfections
  of realistic quantum devices and can be used for thermometry, and  to bound entanglement and
  discord in mixed many-body states. 
\end{abstract}

\maketitle

\prlsection{Introduction}
More than two decades of active research in quantum information processing 
has promoted various quantum technologies, which are believed to 
result in a new industrial revolution \cite{dowling2003quantum}.
One of the major goals, which dates back to Feynmann \cite{feynman1982simulating},
is to simulate complex interacting quantum systems, which are 
intractable with classical computers, 
with an engineered and controllable quantum device, the so-called 
{\it quantum simulator} \cite{cirac2012goals}. 
Unlike general-purpose quantum computers, which are supposed to be programmable to
achieve different tasks, quantum simulators are designed for a specific goal, which
make them easier to realize. Indeed, so far cold atoms \cite{bloch2012quantum} 
and ions \cite{blatt2012quantum} have been used for successfully simulating 
certain tasks. 
Nevertheless, solid state based quantum simulator is still highly in demand due to
the fact that: i) they provide more versatile types of interaction and 
stronger couplings compared to cold atoms and ions;  
ii) the quest towards miniaturization in electronics has reached the quantum 
level, making solid state quantum devices feasible \cite{salfi2016quantum}.

Much theoretical researches have been conducted to understand the highly entangled 
structures appearing in the ground state of quantum many-body systems 
\cite{amico2008entanglement}. 
For a given bipartition $A$ and $B$ of the whole system, 
which is assumed to be in the pure state $\rho_{AB}{=}\ket{\psi_{AB}}\bra{\psi_{AB}}$, 
the {\it entanglement entropy} is quantified by 
$S_\alpha(\rho_A){=}S_\alpha(\rho_B)$, where $\rho_A{=}\Tr_B\rho_{AB}$ and 
$S_\alpha$ is the Renyi entropy, defined as  
\begin{align}
    S_\alpha(\rho) = \frac{1}{1-\alpha}\log \Tr[\rho^\alpha]~,
    \label{Renyi}
\end{align}
for different values of $\alpha$. 
When $\alpha{\to}1$ the Renyi entropy reduces to the von Neumann entropy 
$S_1(\rho) {=} {-}\Tr[\rho\log\rho]$. The importance of the entanglement entropy is 
twofold: i) it quantifies the entanglement between $A$ and $B$; ii) the discovery of 
its {\it area law} dependence in non-critical systems has immensely contributed to 
the development of efficient approximation techniques \cite{eisert2010colloquium}
for describing many-body systems. 
On the other hand, in critical one-dimensional systems with open boundary conditions,  
conformal field theory analysis shows that there is a logarithmic correction, as  
\begin{align} 
S_\alpha(x)=\frac{c}{12}\left(1+\frac{1}{\alpha}\right) \log\left[\frac{2N}{\pi} \sin\left(\frac{\pi x}{N}\right)\right] + \kappa_\alpha
\label{CFT-Uniform}
\end{align}
where $x$ is the size of the contiguous block $A$ starting at one end of system, and 
$N$ is the total size. When $N{\gg}x{\gg}1$ the usual scaling $S_\alpha{\propto} \log x$ 
is obtained. 
This formula is very general and the {\it central charge} 
$c$ only depends on the universality class of the model, while the constants 
$\kappa_\alpha$ are model dependent \cite{calabrese2010parity,calabrese2004entanglement,fagotti2011universal}. 
\begin{figure}[t]
	\centering
	\includegraphics[width=.5\textwidth]{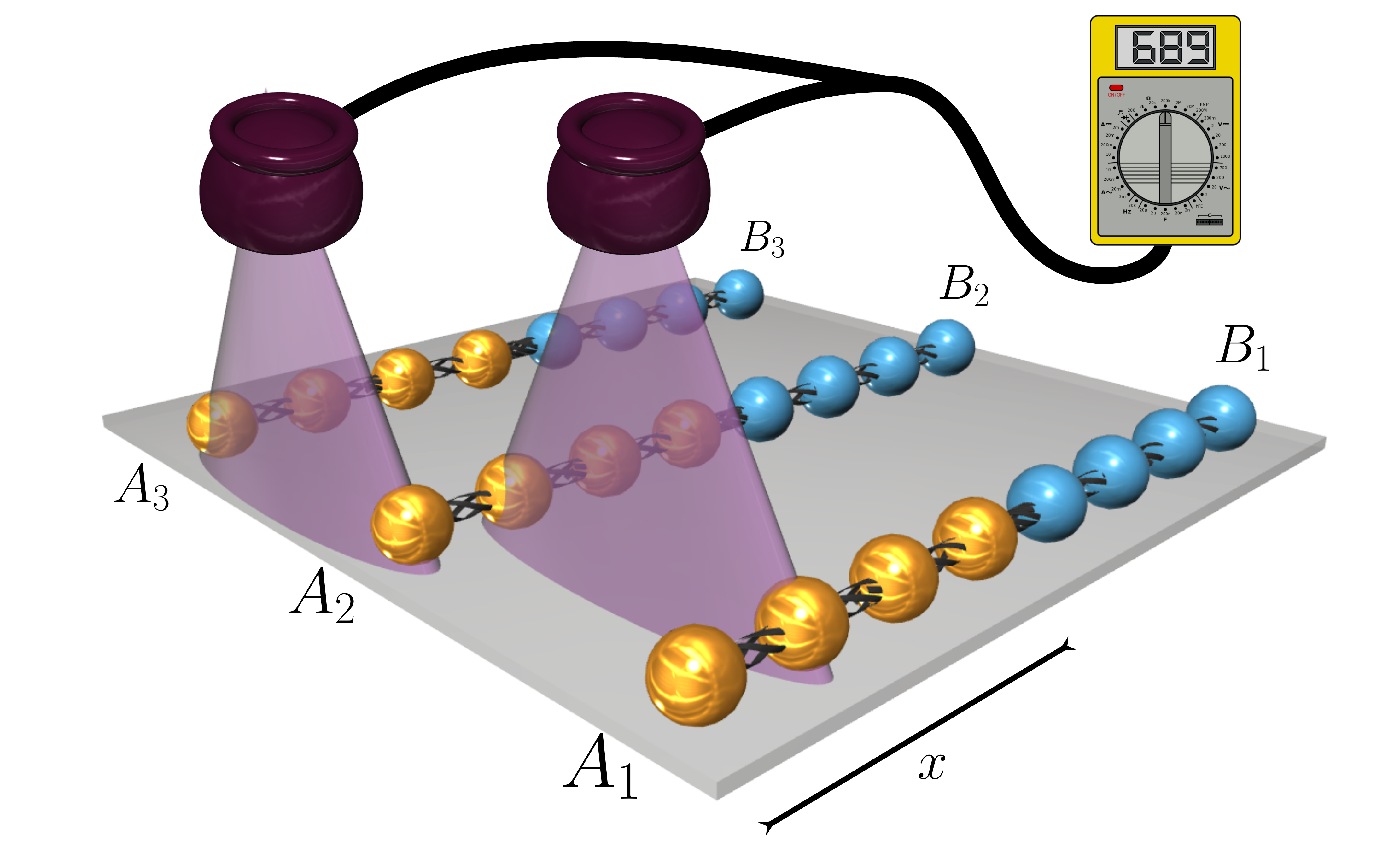}
	\caption{Scheme for the measurement of the Entanglement entropy for
		$\alpha{=}3$ using three copies of the same spin chain. Each spin chain is
		divided into $A_i$ (yellow spins) and $B_i$ (blue spins). By performing
		sequential singlet-triplet measurements of pair of spins in neighbouring 
		chains is it possible to estimate $S_{\alpha{=}3}(\rho_A)$ (see the
		discussion in the text), which measures the entanglement between $A$ and
		$B$. 
	}
	\label{fig:scheme}
\end{figure}
In spite of the extensive theoretical literature on entanglement entropy, 
its experimental measurement is a big challenge.  
For itinerant bosonic particles it has been proposed 
\cite{alves2004multipartite,daley2012measuring}, and recently realized
\cite{kaufman2016quantum}, to use beam splitter operations or discrete 
Fourier transform to measure $S_\alpha$. 
Alternatively measuring entropy through 
quantum shot noise has been proposed \cite{song2012bipartite,klich2009quantum},
but not yet realized. 
On the other hand, in 
non-itinerant spin systems, the situation become even more difficult and
the only proposal so far is to use spin-dependent switches \cite{abanin2012measuring},
which are difficult to build. 

Here we put forward a proposal for measuring $S_\alpha$ in a spin system without 
demanding time-dependent particle delocalition or spin-dependent switches. 
While our setup can be realized in different physical systems, we target it to
solid state systems, such as gated quantum dot chains 
\cite{loss1998quantum,hanson2007spins,baart2016nanosecond,nakajima2016phase,ito2016detection,zajac2016scalable,noiri2016coherent} 
or dopant arrays \cite{kane1998silicon,zwanenburg2013silicon,salfi2016quantum,fuechsle2013using}.  
Our procedure is based on well established singlet-triplet measurements, which are now 
routinely performed either via charge detection \cite{petta2005coherent} 
or capacitive radio-frequency reflectometry \cite{house2015radio,petersson2010charge,frey2012dipole,colless2013dispersive}.

\prlsection{Measuring entanglement entropy}
Our goal is to measure $S_\alpha$ for arbitrarily integer values of $
\alpha{\ge}2$. For simplicity we explain the procedure for $\alpha{=}2$ and 
then generalize it for higher values. Inspired by previous alternative 
proposals 
\cite{horodecki2002method,alves2004multipartite,daley2012measuring,abanin2012measuring,cardy2011measuring}
we make use of two copies of a spin array in the state $\rho_1{\otimes}\rho_2$ (ideally for perfect copies $\rho_1{=}\rho_2$). 
Each copy is identically divided into two complementary 
blocks: $A_1$ and $B_1$ for the first copy,   
$A_2$ and $B_2$ for the second one (see Fig. \ref{fig:scheme}).
Let $x$ be the number of spins in $A_1$ (and $A_2$). 
We define the multi-spin swap operator acting on $A_1$ and $A_2$ as  
\begin{align}
P_{12}^A {\equiv}\bigotimes_{\ell=1}^x{\rm SWAP}({\ell_{A_1},\ell_{A_2}}),
\label{swapP}
\end{align}
where ${\rm SWAP}({\ell_{A_1},\ell_{A_2}})$ swaps the two spins at the $\ell$-th sites 
in $A_1$ and $A_2$. Since all the operators  ${\rm SWAP}({\ell_{A_1},\ell_{A_2}})$   are
commuting it is simple to show that 
\begin{align}
\mean{P_{12}^A}
=\Tr[P_{12}^A\,\rho_1{\otimes}\rho_2]=\Tr[\rho_{A_1}\rho_{A_2}]
=\Tr[\rho_A^2]
\label{S2}
\end{align}
where the last equality holds if the two copies are identical, namely 
$\rho_{A_1}{\equiv}\rho_{A_2}{\equiv}
\rho_A$. Therefore, Eq. \eqref{S2} implies that 
$S_2{=}{-}\log\mean{P_{12}^A}$
can be obtained via a sequential measurement of pairwise swap operators acting 
on the different spins of $A_1$ and $A_2$, as shown in Fig. \ref{fig:scheme}.

The above procedure can be generalized to higher integer values of $\alpha$ 
by considering $\alpha$ copies of the spin array in the state
$\rho^{\otimes \alpha}{=}\bigotimes_{\ell{=}1}^\alpha \rho_\ell$
(where ideally all the $\rho_\ell$'s are equal). 
Remarkably {\it sequential} measurements of multi-spin swap operators acting 
on neighboring copies $a$ and $a{+}1$, namely $P^A_{a,a{+}1}$, 
is sufficient to measure the Renyi entropy. 
This is simple, but not trivial as better explained in the Appendix A, because some $P^A_{(a,a{+}1)}$'s for different $a$ are non-commuting. 
However, we show that the simple sequential measurement, exemplified also in 
Fig. \ref{fig:scheme}, corresponds to the measurement of the operator 
$P^{(A)}_{12\dots\alpha}$ which is defined recursively by the formula
\begin{align}
P^{A}_{12\dots\alpha} = \frac{P^{A}_{\alpha,\alpha{-}1}  P^{A}_{1\dots\alpha{-}1} + 
 P^{A}_{1\dots\alpha{-}1} P^{A}_{\alpha,\alpha{-}1}}{2}.
\end{align}
For example for $\alpha{=}3$ this reduces to 
$P^A_{123}{=}(P^A_{23}P^A_{12} {+} P^A_{12} P^A_{23})/2$ and 
$\mean{P^A_{123}} {=} (\Tr[\rho_{A_1}\rho_{A_2}\rho_{A_3}]
{+} \Tr[\rho_{A_1}\rho_{A_3}\rho_{A_2}])/2$, so that for perfect copies 
$\mean{P^A_{123}}{=} \Tr[\rho_A^3]$. In general using Eq. \eqref{Renyi} 
we have $S_\alpha(\rho_A) {=} (1{-}\alpha)^{-1}\log\mean{P^{A}_{12\dots\alpha}}$. 
We stress that $P^{A}_{12\dots\alpha}$ is ultimately written in terms 
nearest neighbor multi-spin swap operators $P^A_{(a,a{+}1)}$. This makes the 
procedure scalable in the lab as one has to first measure $P^A_{12}$, 
then $P^A_{23}$ and so forth till $P^A_{\alpha{-}1,\alpha}$.

\prlsection{Solid state spin chains} 
When exactly one electron is trapped in each quantum dot, 
the interactions between confined electrons in quantum dot arrays is restricted 
to the spin sector, and is described by the Heisenberg Hamiltonian 
\begin{align} 
H = \sum_{k=1}^{N-1} J_k \boldsymbol{\sigma}_{k}\cdot \boldsymbol{\sigma}_{k+1},
\label{Hamiltonian}
\end{align} 
where $J_k$ is the exchange coupling between neighboring sites and $\boldsymbol{\sigma}_{k}{=}(\sigma^x_k,\sigma^y_k,\sigma^z_k)$ is the vector of Pauli operators acting 
on site $k$. 
The couplings $J_k$ can be locally tuned by appropriately changing the local 
gate voltages. 
The system can be initialized into its ground state either by cooling, 
when temperatures is below its energy gap, or using 
an adiabatic-type evolution \cite{farooq2015adiabatic} when 
temperature is higher.

Singlet-triplet measurements on two electrons trapped in adjacent quantum dots is 
now a well-established technique for spin measurements in solid state physics
\cite{petta2005coherent,petersson2010charge,frey2012dipole,colless2013dispersive,delbecq2016quantum}. 
In a quantum mechanical language the singlet-triplet measurements on a pair of 
electrons in dots $a$ and $b$ correspond to 
{\it projective} measurements of the swap operator, as one can show 
\begin{align}
  {\rm SWAP}(a,b) =  \sum_{\mu=\pm,0} \ket{t_\mu}\bra{t_\mu} - 
  \ket s \bra s = \frac{\openone + \boldsymbol{\sigma}_a\cdot\boldsymbol{\sigma}_b}2,
  \label{swap}
\end{align}
$\ket s {=} (\ket{ {\ua}_a{\da}_b}{-}\ket{
{\da}_a{\ua}_b})/\sqrt 2$ is the singlet state, and 
$\ket{t_+} {=} \ket{ {\ua}_a{\ua}_b}$, $\ket{t_0} {=} (\ket{ {\ua}_a{\da}_b}{+}\ket{
    {\da}_a{\ua}_b})/\sqrt 2$, $\ket{t_{-}} {=} \ket{ {\da}_a{\da}_b}$ are the
  triplet states. 
The outcome of this measurement is either $+1$, for triplet outcomes, and $-1$ for 
the singlet one.

By comparing Eqs. \eqref{swap} and \eqref{swapP} it is now clear that, 
for any given bipartition, we can use a sequence of  
singlet-triplet measurements to obtain the outcome of the 
operators $P^A_{1\dots\alpha}$ and thus compute all the Renyi entropies $S_\alpha$ 
for all integer $\alpha{\ge}2$. As described before, 
and shown also in Fig. \ref{fig:scheme}, the total number of singlet-triplet measurements
to be performed for a single outcome is $x\alpha$ where $x$ is the number of spins 
in subsystem $A$. 
To measure $P^A_{1\dots\alpha}$ we first switch off the $J_k$'s within each array, and 
then lower the barriers between pairs of spins in two different arrays to perform 
the singlet-triplet measurements. 
A recently developed multiplexer structure \cite{puddy2015multiplexed}
containing two parallel arrays of quantum dots is a promising setup, which can be 
adapted for measuring $S_2$ with our proposed mechanism.  
Motivated by this operating device, and for the sake of simplicity, in the rest 
of the paper we focus on $\alpha{=}2$. 
Numerical results are obtained with either Density Matrix Renormalization Group
(DMRG) or exact diagonalization for short chains.

\begin{figure}[t]
	\centering
	\includegraphics[width=.47\textwidth]{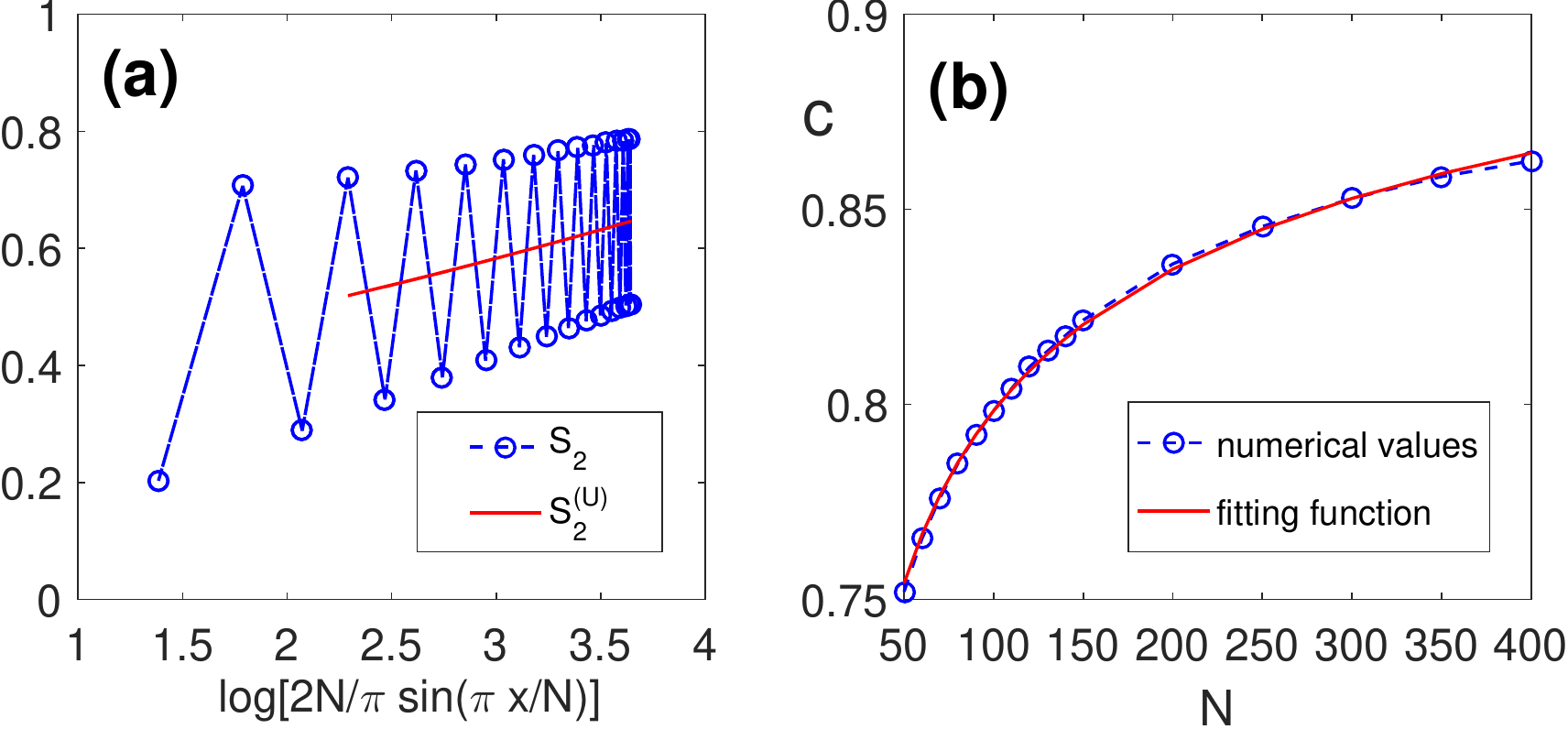}
\caption{(color online) (a) Scaling of $S_2$ and its uniform part $S^{(U)}_2$ in terms 
  of differet sizes of block $A$, for a chain of  $N{=}60$. (b) Scaling of the 
  central charge $c$ as a function of length $N$. }
	\label{fig:uniform}
\end{figure}

\begin{figure*}
	\centering
\includegraphics[width=0.95\textwidth]{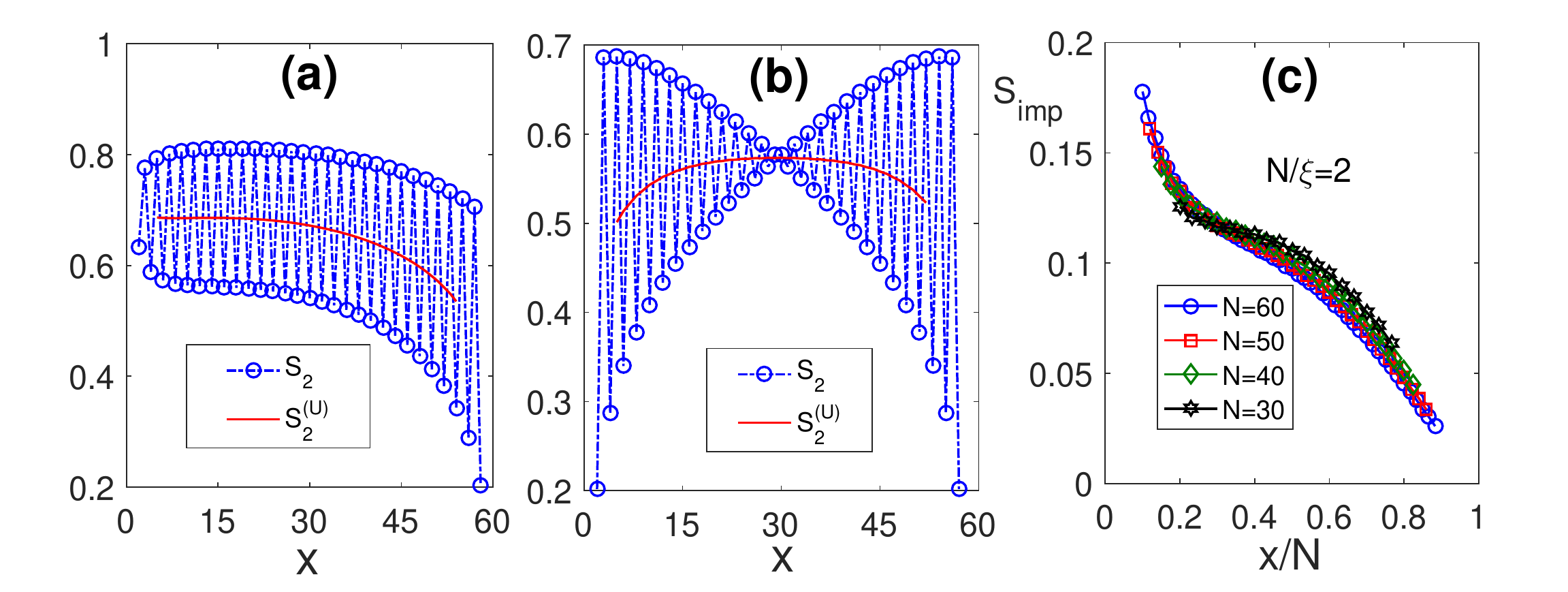}
\caption{(color online) (a) 
  The Renyi entropy $S_2(x,N,\xi)$ and its uniform part $S_2^{(U)}$ in a chain of length $N{=}60$. (b)
  The bulk Renyi entropy $S_2(x{-}1,N{-}1)$ and its uniform part $S_2^{(U)}(x{-}1,N{-}1)$
  in a homogeneous chain of 59 sites without impurity. (c) Data collapse for different 
  chains when $J'$ is tuned to keep $N/\xi{=}2$. 
} 
\label{fig:impurity}
\end{figure*}

\prlsection{Application 1: conformal field theory in the lab}
We first present how field theory predictions, given in Eq. \eqref{CFT-Uniform}, 
can be verified for a uniform chain where $J_k{=}J$, for all $k$'s. 
In the thermodynamic limit $N{\to}\infty$ it is known that the central charge is 
$c{=}1$. 
In Fig.~\ref{fig:uniform}(a) we plot the Renyi entropy $S_2$ as a function of $\log[\frac{2N}{\pi} \sin(\frac{\pi x}{N})]$ in a chain of length $N=60$. 
For open boundary conditions, finite size effects are known 
\cite{calabrese2010parity} to give rise to an 
alternating behaviour of $S_2(x){=}S_2^{(U)}{+}({-}1)^x S_2^{(A)}$. 
Using the methodology of Ref. \cite{sorensen2007quantum} 
we extract the uniform part $S_2^{(U)}$, which is dominant for $N{\to}\infty$ and follows the scaling of 
\eqref{CFT-Uniform}. 
In Fig.~\ref{fig:uniform}(a) we also plot $S_2^{(U)}$ in red colors, showing 
perfect linear scaling. From the slope of this line we can extract the central charge $c$, which asymptocically approaches its thermodynamic limit value, $c{=}1$. This can be
seen in 
Fig.~\ref{fig:uniform}(b) where we also plot the fitting function $c=1-0.7536 N^{-0.2848}$.
Such slow convergence is due to finite-size corrections to the field theory predictions 
\cite{xavier2012finite}, which here, for simplicitly, we have absorbed into the definition
of $c$.

\prlsection{Application 2: impurity entanglement entropy} 
Introducing one or more impurities in the system can change its behaviour drammatiacaly.
A paradigmatic example is the single-impurity Kondo model 
\cite{affleck2008quantum} 
in which a single impurity in a gapless system creates a length scale $\xi$, known
as Kondo length. The scaling features of Kondo physics can be captured by a spin chain
emulation of this model \cite{sorensen2007quantum}. This is described by  
Eq. \eqref{Hamiltonian} where $J_1{=}J'$ while all other 
couplings remain uniform $J_k{=}J$ (for $k {\ge} 2$). 
Moreover, the length scale is determined by $J'$ as $\xi{\propto}e^{{g}/J'}$ 
for some constant $g$. The presence of the impurity modifies the scaling of the
Eq. \eqref{CFT-Uniform} when $x{<}\xi$.  
In order to capture the impurity contribution of the entanglement entropy we extend 
the ansatz of Ref. \cite{sorensen2007quantum} for $S_1$ to generic $S_\alpha$ and
define the impurity entanglement entropy as 
\begin{align} \label{S_imp}
S_\alpha^{(imp)}(x,N,\xi)=S_\alpha^{(U)}(x,N,\xi)-S_\alpha^{(U)}(x-1,N-1),
\end{align} 
where $S_\alpha^{(U)}(x,N,\xi)$ is the Renyi entropy of a block of size $x$ in a chain of length $N$ and impurity coupling $J'$, which determines $\xi$, while
$S_\alpha^{(U)}(x{-}1,N{-}1)$  represents the bulk contribution of the uniform chain when 
the impurity is removed. 
In Fig.~\ref{fig:impurity}(a) we plot the $S_2(x,N,\xi)$ and its uniform part $S_2^{(U)}$ in a chain of length $N=60$. The bulk contribution of the uniform chain, i.e. $S_2(x{-}1,N{-}1)$, and its uniform part $S_2^{(U)}(x{-}1,N{-}1)$ are plotted in Fig.~\ref{fig:impurity}(b). The qualitative difference between Fig.~\ref{fig:impurity}(a) and Fig.~\ref{fig:impurity}(b) 
is due to the different parities (i.e. even and odd) of the chains. 

The emergence of  the length scale implies that $S_2^{(imp)}(x,N,\xi)$ is only a function
of the ratios $S_2^{(imp)}(x/N,N/\xi)$. To verify this scaling we fix $N/\xi$ 
and plot $S_2^{(imp)}$ as a function of $x/N$ for different lengths $N$. To keep
$N/\xi$ fixed $J'$ has to be tuned according to Ref. \cite{bayat2010negativity}.
The results are shown in 
Fig.~\ref{fig:impurity}(c) where, as predicted,  the curves of different chains collapse oneach other. Although the data collapse becomes better by increasing the system size, 
Fig.~\ref{fig:impurity}(c) shows that the scaling predictions 
can be captured even in relatively small chains.

\prlsection{Application 3: entanglement spectrum} 
For any pure state $\rho_{AB}$ the eigenvalues of $\rho_A$ are 
called {\it entanglement spectrum}
\cite{li2008entanglement},
whose analysis is important 
to characterize quantum phase transitions 
\cite{de2012entanglement,bayat2014order}. 
The eigenvalues of $\rho_A$ are the roots of $q(\lambda){=}
\det(\lambda\openone{-}\rho_A)$, which can be written as 
$q(\lambda){=}\sum_k g_k \lambda^k$. According to Ref. \cite{curtright2012galileon}
the coefficients $g_k$ can be obtained algorithmically from $\Tr[\rho_A^\alpha]$
for $\alpha{=}1,\dots,k$. Since these traces can be measured with our procedure one
can build $q(\lambda)$ and hence obtain the full entanglement spectrum. 
Clearly, given a maximum number of copies $\alpha_{\rm max}$, 
one can find the entanglement spectrum for block sizes as large as 
$x{=}\log_2\alpha_{\rm max}$. 

\prlsection{Application 4: thermometry via purity measurement} 
One of the biggest challenges in solid-state experiments is to measure the
{\it true} temperature of electrons, as it is normally higher than 
the temperature of the refrigerator. Remarkably, our scheme enables also to 
measure the electronic temperature via singlet-triplet measurements, assuming 
that the system is in a thermal state $\rho_\beta{=}e^{ {-}\beta H}/Z$. 
Our approach is based on three distinctive features of engineered solid state
structures: i) the exchange integral $J$ 
can be varied; ii) the purity $\mathcal P(\beta,J){=}\Tr[\rho_\beta^2]
{=} e^{ {-}S_2(\rho_\beta)}$ can be measured with our scheme by taking two 
copies and $x{=}N$; 
iii) 
computing the energy expectation 
$E(\beta,J){=}\Tr[H\rho_\beta] $ is reduced to singlet-triplet
measurements on neighboring sites of one of the arrays, thanks to 
Eqs. \eqref{swap} and \eqref{Hamiltonian}. 
A simple calculation reveals that 
$\frac{\partial \mathcal P(\beta,J)}{\partial J}{=} \frac{\beta}J 
[2E(\beta,J){-}E(\beta,2J)] \mathcal P(\beta,J)$. 
Aside from $\beta$ all the quantities in the above equality can be measured
either directly (namely $\mathcal P(\beta,J)$ and $E(\beta,J)$) or 
through the variation of $J$ (namely $\frac{\partial \mathcal P(\beta,J)}{\partial J}$ 
and  $E(\beta,2J)$). 
In summary, thanks to the above equality, 
using different singlet-triplet measurements with different values of $J$ 
it is possible to infer $\beta$ and thus the temperature.

\prlsection{Application 5: bounding entanglement and discord in mixed states}
The Renyi entropy of a block $A$ is a measure of entanglement between $A$ and $B$
only if $\rho_{AB}$ is a pure state. 
However, we show that it is still possible 
to bound the amount of  entanglement and discord 
also for mixed states by measuring both $S_\alpha(\rho_A)$ and 
$S_\alpha(\rho_{AB})$. The distillable entanglement $E_D$, an operational 
entanglement measure, satisfies the 
{\it hashing inequality}
\cite{devetak2005distillation}, 
$E_D {\ge} \max_{X{=}A,B} I_1(X)$, where 
$I_1(X){=}S_1(\rho_X){-}S_1(\rho_{AB})$. 
Similarly, for the quantum discord $D(A|B)$,
which is an asymmetric measure of quantum correlations 
between $A$ and $B$ \cite{ollivier2001quantum}, it is known that 
$D(A|B){\ge}I_1(A)$ and similarly $D(B|A){\ge}I_1(B)$ \cite{fanchini2011conservation}.
The von Neumann entropy $S_1$ can be 
extrapolated \cite{de2015entanglement}
from $S_\alpha$ for different integers $\alpha{\ge}2$, which
can be measured with our scheme. 
However, we show that $I_1$ can also be bounded by directly measuring 
$S_2(\rho_{A})$ and  
$S_2(\rho_{AB})$,
which require only two replicas. Indeed, since
$S_2(\rho){\le}S_1(\rho){\le}f[S_2(\rho)]$, where $f$ is given in Ref. 
\cite{zyczkowski2003renyi}, we obtain $I_1(X){\ge}I_2(X)$ where 
$I_2(X){=}S_2(\rho_X)-f[S_2(\rho_{AB})]$. For either $X{=}A,B$,  $I_2(X)$ thus
provides a measureble lower bound to entanglement and discord. 

\begin{figure}[t]
	\centering
  \includegraphics[width=0.45\textwidth]{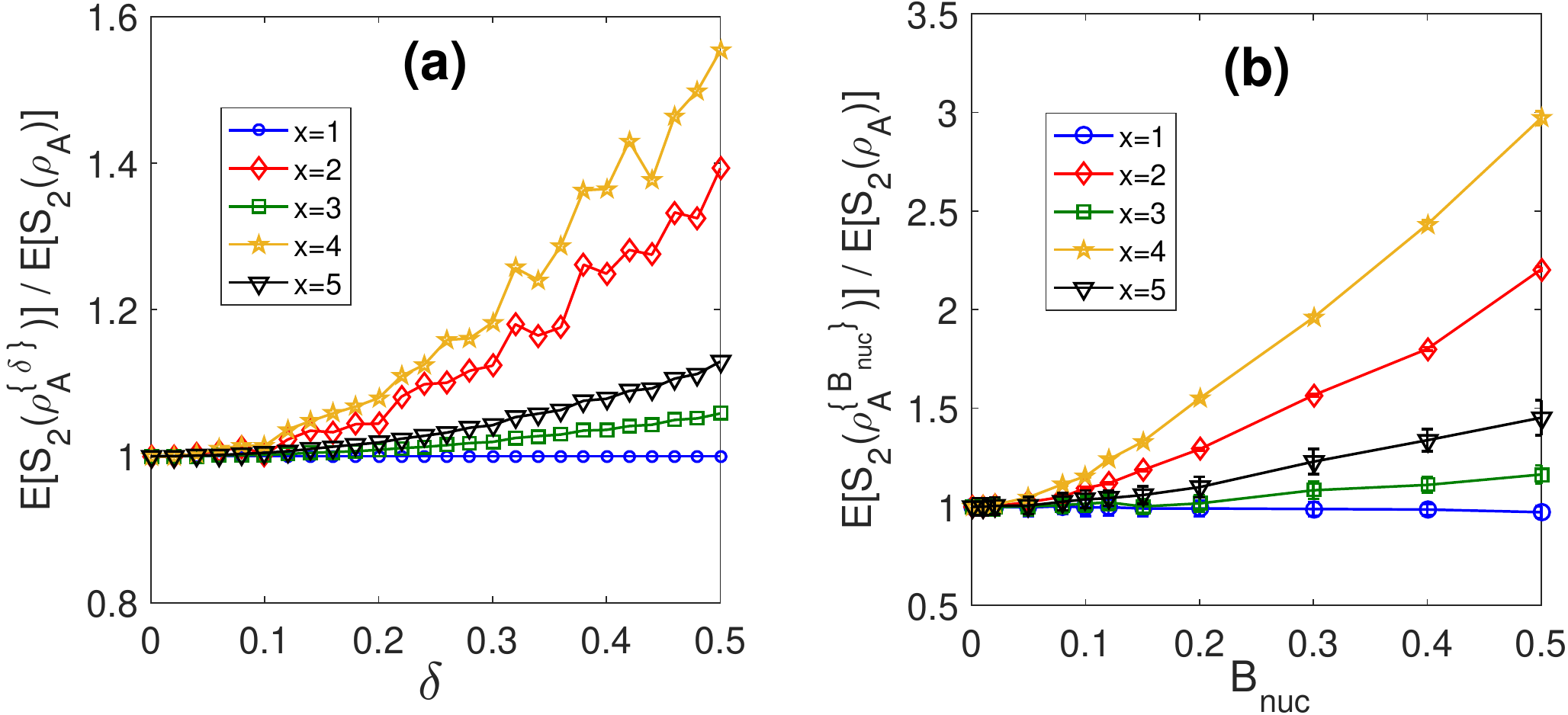}
	\caption{(color online) (a) 
   Average $\mathbb{E}[S_2(\rho^{\{\delta\}}_A)]$ 
over 1000 random sets of couplings $J^{(j)}_k{=}J_k(1{+}\epsilon^{(j)}_k)$,  where
$j{=}1,2$ refers to the different copies, $J_{k}{=}J$ and 
    $\epsilon^{(j)}_k$ is uniformly distributed in $[{-}\delta,{+}\delta]$ 
for different 
block sizes $x$
(b) Monte Carlo simulation of  $\mathbb{E}[S_2(\rho^{\{ B_{\rm nuc}\}}_A)]$  for
the effect of random fields $B_k^{(j)}$ as a function of the
strength of the hyperfine interaction $B_{\rm nuc}$.
In both figures the entropies are normalized with respect to the error-free case 
$S_2(\rho_A)$, and $N{=}10$. 
  }
	\label{fig:imperfect}
\end{figure}

\prlsection{Imperfections}
Realistic experimental imperfections may introduce errors, e.g. 
by making the different copies non-identical. Our protocol provides 
$\Tr[\rho_{A_1}\rho_{A_2}]$, as given in Eq.~\eqref{S2}, which may deviate 
from the ideal case $\Tr[\rho_A^2]$. 
Indeed, imperfect fabrication may 
result in random couplings $J^{(j)}_k{\rightarrow} J_k(1{+}\epsilon^{(j)}_k)$, 
where $j$ is the index of different copies and 
$\epsilon_k^{(j)}$ is a random number uniformly distributed between $[-\delta,+\delta]$.
In Fig.\ref{fig:imperfect}(a) we calculate the average $\mathbb{E}[S_2(\rho^{\{
    \delta\}}_A)]$ 
over 1000 random sets of couplings for different 
block sizes $x$, normalized with respect to its value at $\delta{=}0$. As the figure
shows the average entropy increases by increasing $\delta$. Moreover, up to
$\delta{=}10\%$ the outcomes are almost indistinguishable from the error-free case. 

The second source of imperfections is due to the hyperfine interaction with the 
nuclear spins in the bulk, which effectively introduces an extra term
$\sum_k \boldsymbol{B}^{(j)}_k\cdot \boldsymbol{\sigma}_k$ in the Hamiltonian
\eqref{Hamiltonian}, where each compoenent of the random fields has a normal
distribution with zero mean and variance $B_{\rm nuc}$. 
Unlike the randomness in the couplings, which is constant over different experiments
being due to the fabrication, the effective random fields are different in any
experimental repeatition. To realistically model this, we perform a Monte Carlo 
simulation of real experimental outcomes (see the Appendix C for details). 
The results are shown in Fig.~\ref{fig:imperfect}(b) for different block sizes $x$. 
For realistic values of $B_{\rm nuc}{=}0.1J$ \cite{petta2005coherent} we see that 
the entanglement entropy is only slightly affected by hyperfine interactions. 
The increasing trend of the entropy as a function of the noise, 
which is consistent with Ref. \cite{gatelina}, is further discussed 
in the Appendix B.

\prlsection{Conclusions}
We propose a scheme to expeimentally measure the entanglement between blocks 
 in engineered solid state quantum devices.
Our procedure is based on singlet-triplet measurements which 
are rountinely performed in quantum dot systems.  
All the Renyi entanglement entropies $S_{\alpha}$ for integer $\alpha{\ge}2$
can be measured via $\alpha$ replicas of the system.
Although for uniform chains the convergence of the central charge is slow, 
the logarithmic scaling predictions can already be 
verified with reasonably small system sizes ($\lesssim 60$).  Moreover, in the
Kondo impurity model we found that the impurity contribution in the Renyi entropy 
satisfies a universal scaling law.
Despite the fact this law has been obtained in the thermodynamic limit, remarkably
it can be observed for chains as small as 
$N{\gtrsim} 30$.  In addition, our scheme enables the measurement of 
the purity of the whole system, which allows one to measure the true temperature 
of electrons in a thermal state. 
Our procedure remains effective even in the presence of typical imperfections 
due to imperfect fabrication and hyperfine interactions. 
Although our scheme has been targeted to quantum dot arrays, the same protocol 
can also be realized in other systems, such as dopants in silicon
\cite{zwanenburg2013silicon}.

\begin{acknowledgements}
  \prlsection{Acknowledgements}
 LB and SB have received funding
from the European Research Council under the European
Union’s Seventh Framework Programme (FP/2007-2013) /
ERC Grant Agreement No. 308253. 
AB and SB acknowledge the EPSRC grant EP/K004077/1. 
\end{acknowledgements}

\appendix

\section{Renyi entropies of arbitrary orders via singlet-triplet measurements}

Singlet-triplet (ST) projective measurements can be described by the pair 
of projection operators $\Pi_- {=} \ket s\bra s$, $\Pi_+ {=}\sum_{\alpha{=}{-}1}^1
\ket{t_\alpha}\bra{t_\alpha}$ where $\ket s {=} (\ket{ {\ua}{\da}}{-}\ket{
{\da}{\ua}})/\sqrt 2$ is the singlet state, and 
$\ket{t_1} {=} \ket{ {\ua}{\ua}}$, $\ket{t_0} {=} (\ket{ {\ua}{\da}}{+}\ket{
    {\da}{\ua}})/\sqrt 2$, $\ket{t_{-1}} {=} \ket{ {\da}{\da}}$ are the
triplet states. To relate this measurement with the estimation of the entropy it
is convenient to define the swap operator ${\rm SWAP}$ which is related to the
ST-measurement via $\Pi_\pm{=}(\openone{\pm}{\rm SWAP})/2$. Note that the SWAP
operator can be also written in terms of the Heisenberg spin exchange 
${\rm SWAP}{\equiv}(\openone{+}\sum_{\alpha{=}{x,y,z}}\sigma_\alpha{\otimes}\sigma_\alpha)/2$,
where $\sigma_\alpha$ are the Pauli matrices. 
Since ${\rm SWAP}{=}\Pi_+{-}\Pi_-$ the 
ST-measurement can be understood as a projective measurement of the
SWAP operator: given a two qubit state, this measurement can result in two
different outcomes, either $+1$ or $-1$, with respective probability
$\Tr[\rho\Pi_\pm]$. 

Our strategy to measure the Renyi entropy 
$
    S_\alpha(\rho) {=} (1{-}\alpha)^{-1}\log \Tr[\rho^\alpha],
    $
is based on two fundamental observations:
(i) as shown in \cite{horodecki2002method} if
one has access to $m$ copies of the {\it same} state $\rho$, then $\Tr[\rho^m] {=} 
\Tr[P_{(12\dots m)} \rho^{\otimes m}]$ where $ \rho^{\otimes
m}{=}\rho{\otimes}\rho{\otimes}{\dots}{\otimes}\rho$ and 
$P_{(12\dots m)}\ket{\psi_1}{\otimes}\ket{\psi_2}\dots\ket{\psi_m}{=}
\ket{\psi_m}{\otimes}\ket{\psi_1}\dots\ket{\psi_{m-1}}$ 
is a permutation operator, which swaps different states according to the 
{\it cyclic} permutation $(12\dots m)$; 
(ii) 
the operator $P_{(12\dots m)}$ can be decomposed in terms of a series of swap
between neighboring copies. In the following we show how to interpret these 
multiple swap operations in terms of single-triplet measurements, as shown
in Fig. \ref{fig:scheme}.

We first focus on the measure of the entanglement entropy for
$\alpha{=}2$. This is quite straightforward, but we review this in detail
to introduce the necessary formalism and to clarify why this simple analysis
cannot be applied for $\alpha{>}2$. 
We consider two copies of the same system, each divided into two disjoint
blocks: the first copy is composed by the blocks $A_1$ and $B_1$
and the second one by the blocks $A_2$ and $B_2$. 
From the previous analysis it is clear that 
$S_2{=}\frac12\log\mean{P_{(12)}^A}$, where 
$P_{(12)}^A$ is a multi-spin swap between the spins in $A_1$ and those in $A_2$. 
Let $x$ be the number of spins in $A_1$ (and $A_2$) and let 
$i_1\dots i_x$ be the indices of the spin in $A_1$, while
$j_1,\dots j_x$ are the indices of the spins in $A_2$. 
Blocks $A_1$ and $A_2$ are non-overlapping. 
The multi-spin swap then reads $P_{(12)}^A
{\equiv}\bigotimes_{\ell=1}^x{\rm SWAP}_{i_\ell
j_\ell}$
where ${\rm SWAP}_{ij}$ is the swap operator between the pair of spins $(i,j)$. 
Therefore, in order to measure the expectation value $\mean{P_{(12)}^A}$, 
one has to perform a series of multiple singlet-triplet measurements between
different pairs $(i_\ell,j_\ell)$ of spins, and collect the resulting
statistics. Indeed, the first projective measurement will result in an outcome
$\pm 1$ and a collapse of $\rho_{A_1A_2B_1B_2}$ into the (non-normalized) state 
$\Pi_\pm^{(i_1,j_1)}\rho_{A_1A_2B_1B_2} \Pi_\pm^{(i_1,j_1)}$, where $\Pi_\pm^{(i_1,j_1)}$ is
the ST projector for the pair $(i_1,j_1)$; after the $x$ projective
measurements between all pair of spins the outcome will be $(\prod_{\ell{=}1}^x
\beta_\ell) 1$, where $\beta_\ell{=}\pm$, with probability 
$\Tr[\rho_{A_1A_2B_1B_2} \bigotimes_{\ell{=}1}^x \Pi_{\beta_\ell}^{(i_\ell,j_\ell)}]$.
Therefore, running these projections many-times will enable an experimental
evaluation of  $\mean{P_{(12)}^A}$.

We now consider the case $\alpha{=}3$. 
For convenience we write $P_{(12)}^A {=} \Pi_{(12)}^+ {-}\Pi^-_{(12)}$ 
in terms of the projection
operators (indeed, as shown before also $P_{(12)}^A$ has eigenvalues $\pm1$).
We first perform a sequential set of ST-measurements on copies $(1,2)$, with
outcome $\beta_{1}$ and then
do the same measurement on copies $(2,3)$, with outcome $\beta_2$. 
We introduce the notation $P^A_{(23)}\circ P^A_{(12)}$ to describe this process. 
After the first measurement, the (non-normalized) state of the system will be 
$\Pi_{(12)}^{\beta_1} \rho \Pi^{\beta_1}_{(12)}$, 
where $\rho{=}\rho_1{\otimes}\rho_2{\otimes}\rho_3$, 
while after the two sets of
measurements it is 
$\Pi_{(23)}^{\beta_2} \Pi_{(12)}^{\beta_1} \rho \Pi^{\beta_1}_{(12)}\Pi_{(23)}^{\beta_2} $.
Therefore, 
\begin{align}
  \mean{P^A_{(23)}\circ P^A_{(12)} } &= \sum_{\beta_2}\sum_{\beta_1} \beta_1 \beta_2 \Tr\left[
  \Pi_{(23)}^{\beta_2} \Pi_{(12)}^{\beta_1} \rho \Pi^{\beta_1}_{(12)}\Pi_{(23)}^{\beta_2}\right]
\nonumber
\\&
  = \sum_{\beta_2}\sum_{\beta_1} \beta_1 \beta_2 \Tr\left[\Pi^{\beta_1}_{(12)}
  \Pi_{(23)}^{\beta_2} \Pi_{(12)}^{\beta_1} \rho \right]
\nonumber
\\&
  = \sum_{\beta_1} \beta_1 \Tr\left[\Pi^{\beta_1}_{(12)}
  P_{(23)}^{A} \Pi_{(12)}^{\beta_1} \rho \right]
\nonumber \\&
  = \frac12\left(\Tr[P^{A}_{(12)}
  P_{(23)}^{A} \rho ] + 
   \Tr[P_{(23)}^{A} P_{(12)}^{A} \rho ]\right)
\nonumber \\&
  = \frac12\left(\Tr[P^{A}_{(123)}\rho ] + \Tr[P_{(132)}^{A} \rho ]\right)
\nonumber \\&
  = \frac12\left(\Tr[\rho_1\rho_2\rho_3] + \Tr[\rho_1\rho_3\rho_2]\right),
  \label{s3}
\end{align}
where we used multiple times the fact that $\Pi^{\pm}_{(ab)} {=} 
(\openone{\pm}{P}^A_{(ab)})/2$
and, in the last equation, that $P^A_{(123)}$ and $P^A_{(132)}$ are permutation 
operators, and $(123),(132)$ different cycles. 
The above equation shows that, because of the non-commutative nature of 
$P_{(12)}^A$ and $P_{(23)}^A$, the sequential process described in Fig. \ref{fig:scheme} 
ends up in the measurement of a combination of different permutation operators.

We now generalize the above argument for higher values of $\alpha$. 
We apply sequential ST-measurements on neighbouring copies, using the notation 
$P_{(12)}^A{\circ} P_{(23)}^A{\circ}{\cdots}{\circ} P_{(\alpha{-}1,\alpha)}^A$, meaning that we first
perform $P_{(12)}^A$ and so forth. As already seen for $\alpha{=}3$, 
the reason for this notation is that, as we show, 
$\mean{P_{(12)}^A{\circ} P_{(23)}^A{\circ} \cdots{\circ} P_{(\alpha{-}1,\alpha)}^A}{\neq}
\mean{P_{(12)}^AP_{(23)}^A\cdots P_{(\alpha{-}1,\alpha)}^A}$. Indeed, after the 
first measurement of $P_{(12)}^A$, with outcome $\beta_1{=}\pm$, the
(non-normalized) state is $\Pi_{(12)}^{\beta_1}\rho_{AB}\Pi_{(12)}^{\beta_1}$.
At later stages one performs sequentially the other measurements $P_{(j,j{+}1)}^A$,
getting the outcomes
$\beta_j$. Taking the averages one then finds that
$$\mean{P_{(\alpha{-}1,\alpha)}^A{\circ}{\cdots}{\circ} P_{(12)}^A}{=} \Tr[\mathcal
    P_{\alpha{-}1,\alpha}[\cdots\mathcal
        P_{23}[\mathcal P_{12}[\rho] ]]\cdots],
$$ where 
$\mathcal P_{j,j{+}1}[\rho]{=} \sum_{\beta_j} \beta_j
\Pi_{(j,j{+}1)}^{\beta_j}\rho\Pi_{(j,j{+}1)}^{\beta_j}$.
Using the cyclic property of the trace and the identity 
$\mathcal P_{j{-}1,j}[P^A_{(nab\dots)}]= [
P^A_{(j{-}1,j,a,\dots)}{+} P^A_{(j,j{-}1,a,\dots)}]/2 $
multiple times (where $a{>}j$, $b{>}j$ and so forth), one finds that 
$\mean{P_{(\alpha{-}1,\alpha)}^A{\circ}{\cdots}{\circ} P_{(12)}^A} {=}2^{2{-}\alpha}
\sum_c \mean{P^A_{c}}$ where $c$ are $2^{\alpha{-}2}$ different {\it cycles}, 
namely cyclic permutations of the elements $1,\dots,{\alpha}$. For instance, for
$\alpha{=}3$ one has $c{=}\{(123),(132)\}$. 
From the above expression it turns out that, if the copies are perfect, then the
different cycles have the same expectation value 
$\mean{P^A_{c}}{=}\Tr[\rho^\alpha]$ and therefore
\begin{align}
S_\alpha=\frac{1}{1{-}\alpha}\log\mean{P_{(\alpha{-}1,\alpha)}^A{\circ}{\cdots}{\circ}
P_{(12)}^A}~.
\label{SRenyiMeasurement}
\end{align}
On the other hand, if the different copies are not exactly equal, then there may
be an extra error (see the numerical examples in the main text). 
In Eq.\eqref{SRenyiMeasurement} each $P_{(j,j{+}1)}^A$ 
requires $x$ ST-measurements ($x$ being the number of spins in $A_i$), so the
total number of ST-measurements for a single outcome is $x\alpha$. 

\section{Imperfections}

\begin{figure}[t]
  \centering
  \includegraphics[width=0.45\textwidth]{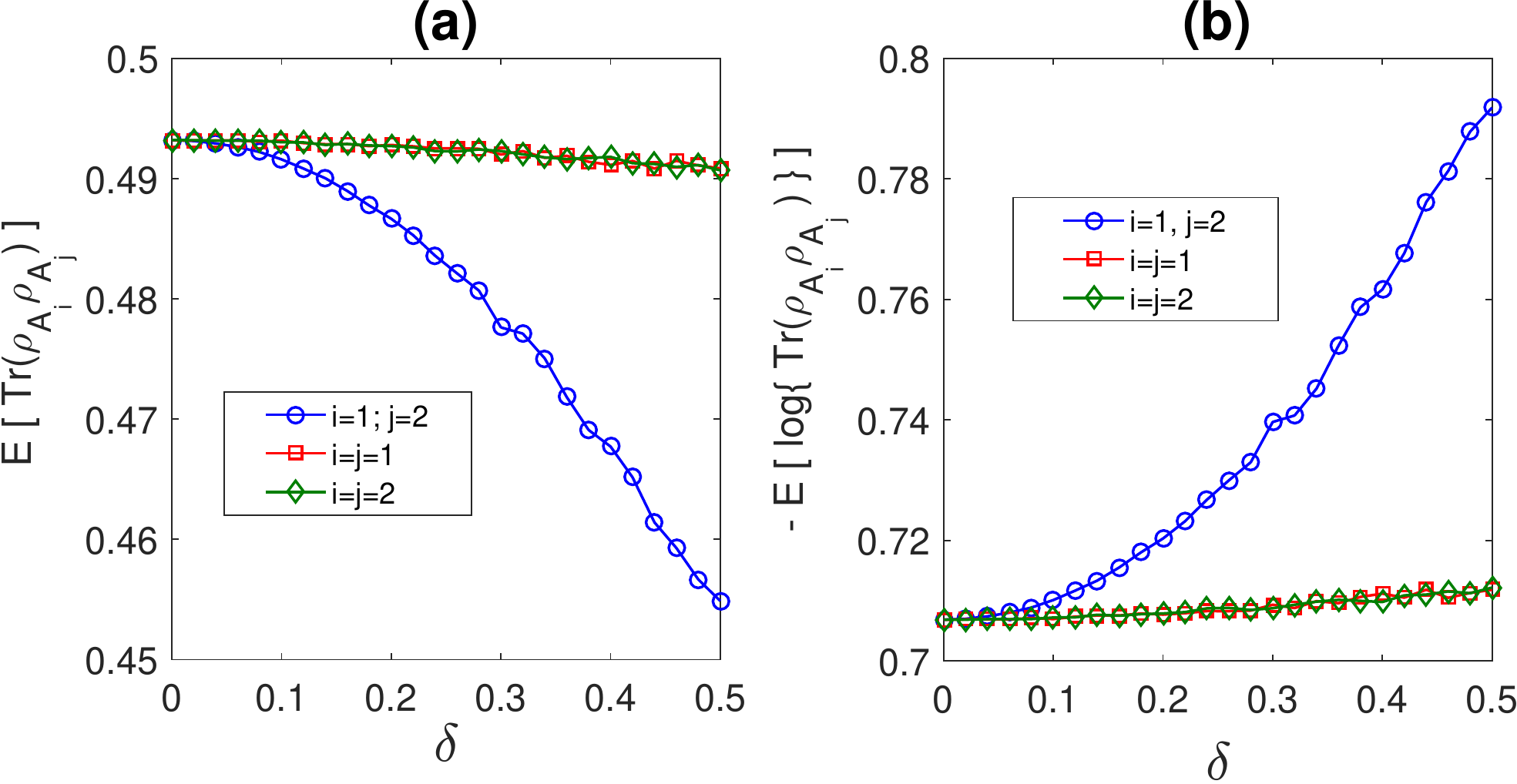}
  \caption{(a) Average purity $ \mathbb{E}[\Tr[\rho_{A_i}\rho_{A_j}]]$
    and (b) average Renyi entropy $ \mathbb{E}[{-}\log\Tr[\rho_{A_i}\rho_{A_j}]]$ 
    as a function of the noise strength 
  $\delta$, calculated with our strategy (where $i{\neq} j$ and each copy feels a different noise) and with the standard procedure (where $i{=}j$). In all cases $N{=}10$, $x{=}5$. }
  \label{imperf}
\end{figure}
We compare the role of the imperfections in our protocol and in the standard 
one in Fig. \ref{imperf}. We consider the noise model discussed in the main text 
where the couplings $J_k$ are different over the different copies, namely
$J_k^{(j)} {=} J_k (1{+}\epsilon_k^{(j)})$ where $j$ refers to the index
of the copies. The errors $\epsilon_k^{(j)}$ are independent identically 
distributed according
to a uniform distribution with zero mean and variance $\delta$. For simplicity we consider 
only $\alpha{=}2$.  
In Fig. \ref{imperf}(a) we show that the average purity estimated with 
our strategy, i.e.  
$\mathbb{E}[\Tr[\rho_{A_1}\rho_{A_2}]]$,  is smaller than  the 
one evaluated with the standard procedure on each chain, i.e. 
$ \mathbb{E}[\Tr[\rho_{A_1}^2]]$ and 
$ \mathbb{E}[\Tr[\rho_{A_2}^2]]$. On the other hand, since the 
Renyi entropy is minus the logarithm of the purity, the average Renyi 
entropy displays the 
opposite behavior, as shown in Fig. \ref{imperf}(b). 
 This can be explained
with the following argument. When $\epsilon_k^{(j)}{\ll}1$ we can expand the states
$\rho_{A_j}$ in series of $\epsilon_k^{(j)}$ and write
\begin{align}
  \rho_{A_j} \simeq \rho_A + \sum_k \epsilon_k^{(j)} \rho'_k + 
  \sum_{k,\ell} \epsilon_k^{(j)} \epsilon_\ell^{(j)} \rho''_{k,\ell},
\end{align}
where $\rho'_k$ and $\rho''_{k,\ell}$ are the first order and second-order expansion, 
which are independent on the index $j$. 
Therefore, 
\begin{align}
  \mathbb{E}[\Tr[\rho_{A_1}\rho_{A_2}]] \simeq \Tr[\rho_A^2] + 2\delta^2\sum_k \Tr[\rho''_{k,k} \rho_A]~,
  \label{tr}
\end{align}
while
\begin{align}
  \label{tr2}
  \mathbb{E}[\Tr[\rho_{A_1}^2]] &= \mathbb{E}[\Tr[\rho_{A_2}^2]] \\&\simeq \Tr[\rho_A^2] + 
  \delta^2\sum_k \left(2\Tr[\rho''_{k,k} \rho_A] + \Tr[\rho'_k{}^2]\right)~.
  \nonumber
\end{align}
The results of 
Fig. \ref{imperf} show that the quantity  $\sum_k\Tr[\rho''_{k,k} \rho_A]$ is negative
and lowers the average purity. On the other hand, $\Tr[\rho'_k{}^2]$ is clearly
positive, being the trace of the square of an operator. 
The latter term then partially removes the effect of the negative one and 
makes $\mathbb{E}[\Tr[\rho_{A_i}^2]] 
{\approx} \Tr[\rho_A^2]$. On the other hand, independent copies 
are slightly more affected by the uncorrelated noise and $\mathbb{E}[\Tr[\rho_{A_1}\rho_{A_2}]]{<}
\mathbb{E}[\Tr[\rho_{A_i}^2]]{\lesssim } \Tr[\rho_A^2]$. The 
error is nonetheless small, since it is  smaller than 10\% even for 50\% of randomness.

\section{Monte Carlo simulation of random fields} 
As explained in the main text, hyperfine interactions with the nuclear spins result 
in the couplings $\sum_k \boldsymbol{B}_k\cdot \boldsymbol{\sigma}$. The random 
effective fields $\boldsymbol{B}_k$ 
are assumed  to be constant after each different ST-measurement required to get 
a single outcome. However, to get the necessary statistics to estimate 
\eqref{SRenyiMeasurement} one has to repeat the experiment many times. Since each 
time the system is re-initialized, the corresponding random fields may be different. 
To study this kind of imperfections we perform the following Monte Carlo simulation
\begin{enumerate}
  \item We generate the random fields $B^{\alpha}_{k,c}$ for different $\alpha{=}x,y,z$, 
    different sites $k{=}1,\dots,N$ and different copies $c{=}1,\dots,\alpha$ independently
    according to a Gaussian distribution, with zero mean and variance $\sigma$. 
  \item We calculate, with exact numerical diagonalization, 
    the quantum mechanical probability 
    $$p{=}\sum_{\prod_j \beta_j = {+1}} \Tr\left[\prod_j 
    \Pi^{\beta_j}_{(j,j{+}1)} \rho \prod_l \Pi^{\beta_l}_{(l,l{+}1)}\right],$$ 
    to get the outcome ${+}1$. In the above equation $\rho{=}\rho_1{\otimes}\cdots{\otimes}
    \rho_\alpha$ where $\rho_j$ is the ground state of the $j$-th copy with the random fields. 
    In general therefore $\rho_i{\neq}\rho_j$. 
  \item We generate a random number q in $[0,1]$. If $q{<}p$ we say that the outcome of the 
    sequence of projective measurements is ${+}1$, otherwise it is ${-}1$. 
  \item We repeat the steps 1,2,3 $T$ times to estimate 
    $\mean{P_{(\alpha{-}1,\alpha)}^A{\circ}{\cdots}{\circ}P_{(12)}^A}$, and then calculate
    $S_\alpha$. 
  \item We repeat the steps 1,2,3,4 $K$ times to calculate the average $\mathbb{E}[S_\alpha]$ and
    the variance. 
\end{enumerate}

In the following figures we show the outcome of this procedure for $N{=}10$, $x{=}1,\dots,N/2$ 
as a function of $\sigma$, when $K{=}T{=}100$. 
The red lines correspond to the exact 
numerical calculation of $S_\alpha$ when $\sigma{=}0$. 

\begin{center}
\includegraphics[width=.45\textwidth]{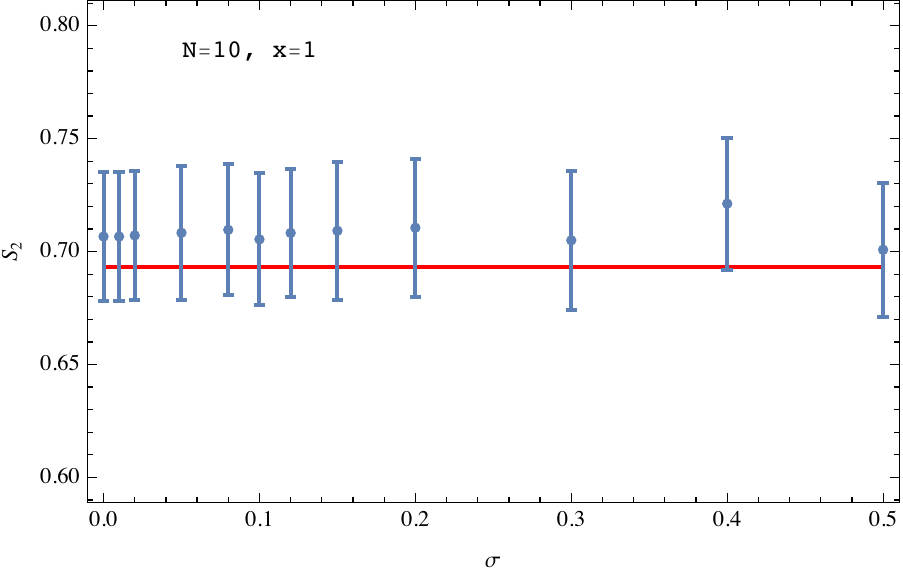}

\includegraphics[width=.45\textwidth]{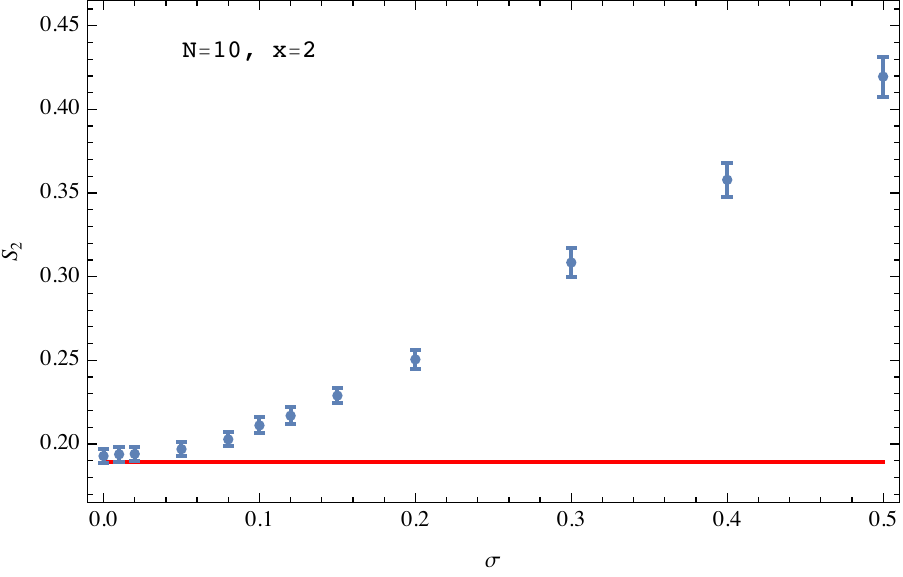}

\includegraphics[width=.45\textwidth]{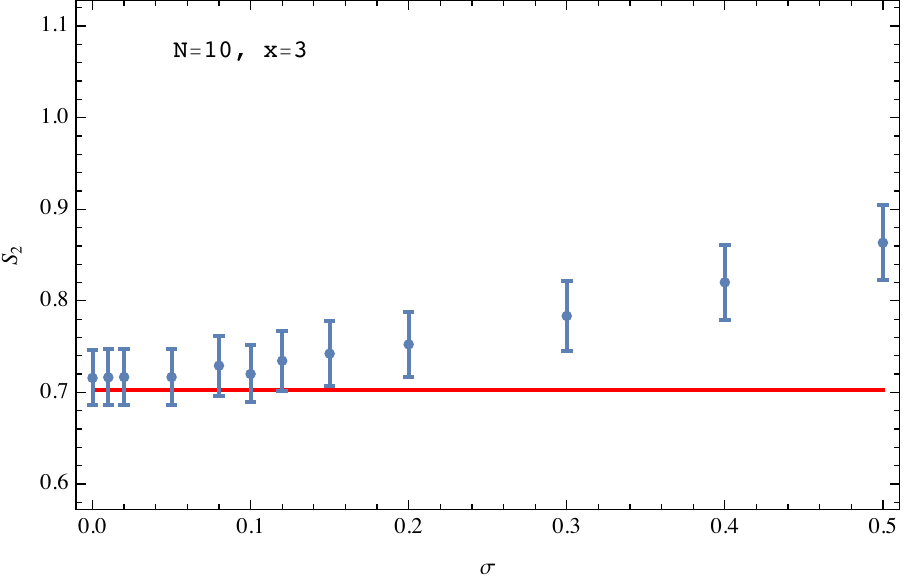}

\includegraphics[width=.45\textwidth]{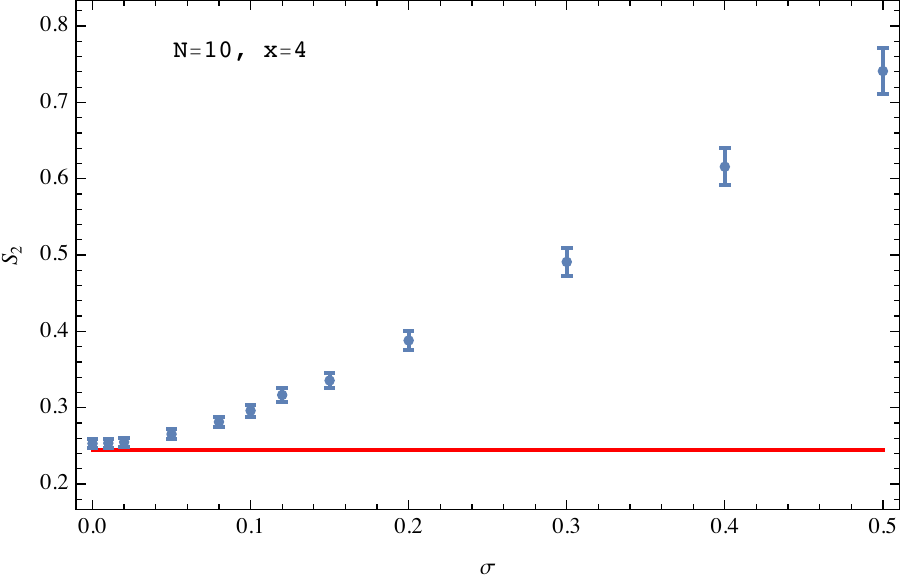}

\includegraphics[width=.45\textwidth]{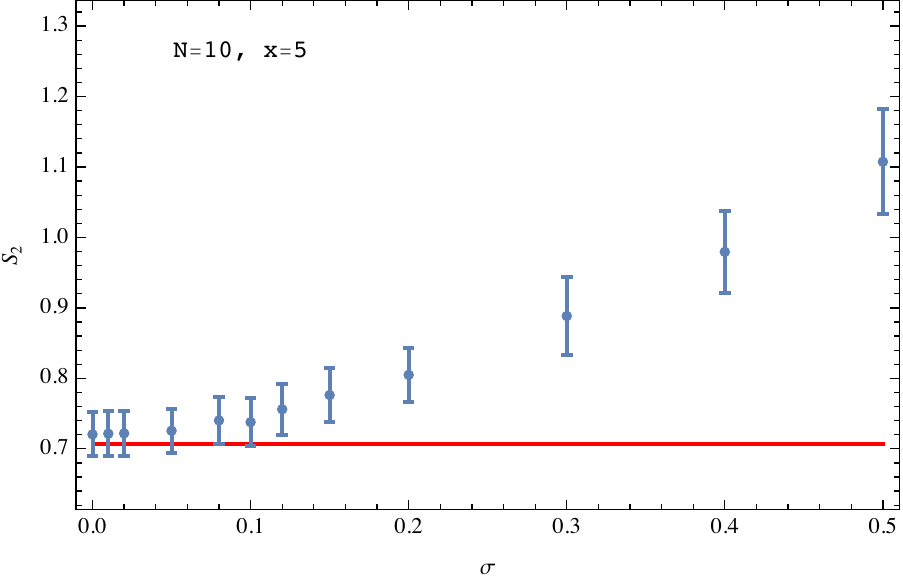}
\end{center}


%
\end{document}